\documentclass[twocolumn,prc,superscriptaddress,amsmath,amssymb]{revtex4-2}

\usepackage{amsmath,amsthm,amssymb}
\usepackage{graphicx}
\usepackage{color}
\usepackage{xcolor}
\usepackage{float}
\usepackage{multirow,array}
\usepackage[colorlinks,
bookmarks=true,
linkcolor=blue,
urlcolor=blue,
anchorcolor=black,
citecolor=blue
]{hyperref}

\colorlet{darkblue}{blue!60!black}

\begin{document}

\title{Baryon density dependence of viscosities of the quark-gluon plasma at hadronization}

\author{Zhidong Yang}
\affiliation{School of Physics and Astronomy, Shanghai Key Laboratory for Particle Physics and Cosmology, and Key Laboratory for Particle Astrophysics and Cosmology (MOE), Shanghai Jiao Tong University, Shanghai 200240, China}


\author{Yifeng Sun}
\affiliation{School of Physics and Astronomy, Shanghai Key Laboratory for Particle Physics and Cosmology, and Key Laboratory for Particle Astrophysics and Cosmology (MOE), Shanghai Jiao Tong University, Shanghai 200240, China}

\author{Lie-Wen Chen}
\thanks{Corresponding author: lwchen@sjtu.edu.cn}

\affiliation{School of Physics and Astronomy, Shanghai Key Laboratory for Particle Physics and Cosmology, and Key Laboratory for Particle Astrophysics and Cosmology (MOE), Shanghai Jiao Tong University, Shanghai 200240, China}

\date{\today}

\begin{abstract}
{The $\phi$ meson and $\Omega$ baryon provide unique probes of the properties of the quark-gluon plasma (QGP) at hadronization in relativistic heavy-ion collisions. Using the quark recombination model with the quark phase-space information parameterized in a viscous blastwave, we perform Bayesian inference of the shear and bulk viscosities of the QGP at hadronization with a temperature of $T\sim 160$ MeV by analyzing the $\phi$ and $\Omega$ data in Au+Au collisions at $\sqrt{s_{\rm NN}}=$ 19.6-200~GeV and Pb+Pb collisions at $\sqrt{s_{\rm NN}}=$ 2.76~TeV, corresponding to a baryon chemical potential variation from $\mu_B\approx 0$ (at $\sqrt{s_{\rm NN}}= 2.76$~TeV) to $200$ MeV~(at $\sqrt{s_{\rm NN}}= 19.6$ GeV). We find that the shear viscosity to enthalpy ratio $\eta T/(\epsilon +P)$ of the QGP at hadronization decreases as $\mu_B$ increases, with $\eta T/(\epsilon +P)\approx 0.18$ at $\mu_B=0$ and $\eta T/(\epsilon +P)\approx 0.08$ at $\mu_B=200$ MeV, while the corresponding specific bulk viscosity is essentially constant with $\zeta T/(\epsilon + P)=0.02\sim 0.04$ for $\mu_B<200$ MeV.
Our results suggest that the QGP at hadronization ($T\sim 160$ MeV) with finite baryon density is more close to perfect fluid than that with zero baryon density.}

\end{abstract}


\maketitle

\section{Introduction}

Lattice Quantum Chromodynamics (QCD) calculations predict a transition from ordinary hadronic matter to a new state of matter that consists of deconfined quarks and gluons, called quark gluon plasma (QGP) \cite{Fodor:2004nz}. The QGP is believed to have existed in the early universe $10^{-6}s$ after the big bang and can be created in relativistic heavy-ion collisions (HICs) at the BNL Relativistic Heavy Ion Collider (RHIC) and the CERN Large Hadron Collider (LHC). Exploring the QCD phase diagram as well as the transport properties of the QGP is one of the most fundamental problems in high-energy nuclear physics (see, e.g., Ref.~\cite{Shen:2020gef}). It is known that at zero baryon density or baryon chemical potential ($\mu_B=0$), the transition between QGP and hadronic matter is a smooth crossover~\cite{Bhattacharya:2014ara,Borsanyi:2020fev}. Further calculations from lattice QCD suggest that the crossover line extends up to $\mu_B\sim$ 250-300 MeV \cite{Borsanyi:2021sxv}. However, it is not yet known if there exists a critical point where the crossover transforms into a first-order phase transition at higher baryon densities. In fact, the main goal of the beam energy scan (BES) program at RHIC is to investigate the phase diagram of QCD and locate the critical point \cite{An:2021wof}.

One of the most significant discoveries made at RHIC and LHC is that the QGP behaves like a near-perfect fluid, characterized by an exceptionally small shear viscosity to entropy density ratio $\eta/s$, close to the universal lower bound $1/4\pi$ based on the anti-de Sitter/conformal field theory (AdS/CFT) correspondence~\cite{Kovtun:2004de}. This is surprising since the QGP was initially expected to be a weakly interacting gas of quarks and gluons but turned out to be a strongly coupled fluid. This discovery has attracted intense interest in the transport properties of the QGP fluid, which are closely related to the underlying strong interactions between quarks and gluons. In recent years, theoretical calculations on the shear ($\eta$) and bulk ($\zeta$) viscosity have been extensively explored~\cite{Meyer:2007ic,Meyer:2007dy,Meyer:2009jp,Mages:2015rea,Astrakhantsev:2017nrs,Astrakhantsev:2018oue,Borsanyi:2018srz,Christiansen:2014ypa,Liu:2016ysz,Ghiglieri:2018dib,Rougemont:2017tlu,Grefa:2022sav}, usually considering $\mu_B=0$ and focusing on the temperature dependence of viscosities.

The viscosities of the QGP have significant effects on the final observables of HICs~\cite{Teaney:2003kp,Ryu:2015vwa}, allowing us to constrain their values with experimental data. Early studies employing viscous hydrodynamics usually assumed a constant $\eta/s$ over the entire evolution and found $\eta/s=0.08\sim 0.2$~\cite{Romatschke:2007mq,Schenke:2010rr,Gale:2012rq,Niemi:2015qia}. Recent researches using Bayesian statistical analysis with multi-stage models that integrate initial conditions, viscous hydrodynamics and hadronic transport
obtained constraints on the temperature dependence of the shear and bulk viscosities of the baryon-free QGP~\cite{Bernhard:2015hxa,Bernhard:2016tnd,Bernhard:2019bmu,JETSCAPE:2020shq,JETSCAPE:2020mzn,Nijs:2020ors,Nijs:2020roc,Nijs:2022rme,Parkkila:2021tqq,Parkkila:2021yha,Heffernan:2023utr}. Overall, most recent analyses have yielded consistent results and found $\eta/s \approx 0.16$ at at pseudo-critical temperature $T_{\rm pc}\approx 160$~MeV for the baryon-free QGP~\cite{Nijs:2022rme,Heffernan:2023utr}.

Studies have also found that the viscosities of QCD matter depend on baryon chemical potential~\cite{Denicol:2013nua,Kadam:2015xsa,Rougemont:2017tlu,Soloveva:2020hpr,McLaughlin:2021dph,Grefa:2022sav}. By assuming a constant value of $\eta/s$ for each collision energy, hybrid models that incorporate hadronic transport and viscous hydrodynamics show that different values of effective shear viscosity are required to describe the data at different collision energies~\cite{Karpenko:2015xea,Auvinen:2017fjw}.
The baryon chemical potential dependence of $\eta/s$ is explored in Refs.\ \cite{Shen:2020jwv,Gotz:2022naz,Shen:2023awv}. In general, due to large uncertainties in the initial conditions and equation of state, hydrodynamics simulations at finite $\mu_B$ are under development~\cite{Akamatsu:2018olk,Du:2019obx,Shen:2020jwv,Wu:2021fjf,Shen:2023awv} and a quantitative estimate for QGP's viscosities at finite $\mu_B$ is still challenging.

In this work, we present a novel approach to constrain the shear and bulk viscosities of QGP at finite $\mu_B$ with a temperature of $T\sim 160$ MeV by using the $\phi$ meson and $\Omega$ baryon observables in relativistic HICs from LHC to RHIC-BES energies.
We find
that the QGP at $T\sim 160$ MeV with finite baryon density is more close to perfect fluid than that with zero baryon density.
The similar approach has been recently applied to explore the viscosities of the baryon-free QGP at hadronization~\cite{Yang:2022ixy}.
In our approach, hadrons are produced through quark recombination~\cite{Greco:2003xt,Greco:2003mm,Fries:2003vb,Fries:2003kq,Hwa:2002tu,Chen:2004dv,Kolb:2004gi,Fries:2008hs,He:2010vw,Zhao:2020wcd} with the phase-space distribution of quarks at hadronization parameterized in a viscous blastwave~\cite{Teaney:2003kp,Jaiswal:2015saa,Yang:2016rnw,Yang:2020oig,Yang:2022yxa} which includes non-equilibrium deformations of thermal distributions due to shear and bulk stresses. The viscous effects for the QGP at hadronization are then imported into $\phi$ and $\Omega$ through the recombination process.
Since the $\phi$ and $\Omega$ have relatively small hadronic interaction cross sections~\cite{Shor:1984ui}, they thus carry direct information of QGP at hadronization with negligible hadronic effects~\cite{Shor:1984ui,VanHecke:1998yu,Chen:2006vc,Chen:2008vr,Auvinen:2016uuv,Hwa:2016qtb,Ye:2017ooc,Pu:2018eei,Song:2019sez}.
It should be pointed out that blastwave models obtain their parameters through fits to data and are independent of initial conditions and equation of state, thus providing a complementary way to hydrodynamic simulations~\cite{Yang:2016rnw,Yang:2020oig}.

The paper is organized as follows. In Section~\ref{sec:qr}, we introduce the models and methods employed in our present work. In Section~\ref{sec:results}, we present the results and discussions, and the conclusion is given in Section~\ref{sec:conclusion}.

\section{Model and Methods}
\label{sec:qr}

\subsection{Theoretical model}

Quark recombination or coalescence models were initially proposed to explain the baryon-over-meson enhancement and valence quark number scaling observed in RHIC Au+Au collisions~\cite{Greco:2003xt,Greco:2003mm,Fries:2003vb,Fries:2003kq,Hwa:2002tu,Chen:2004dv,Kolb:2004gi}.
In a recent work \cite{Yang:2022ixy}, we introduce viscous corrections into quark recombination, and here give a brief overview of the formalism. In the following $\mathbf{r}$ is the 3-space position, $\mathbf{p}$ is the 3-momentum and $m$ is particle mass. The 4-momentum of hadrons are denoted as $p^\mu = (E,~\mathbf{p})$ with $E=\sqrt{m^2+p^2}$. Following Refs.~\cite{Fries:2003kq,Fries:2003vb,Fries:2008hs}, the momentum distribution of mesons is given by
\begin{eqnarray}
	E\frac{dN_M}{d^3\bf p}&=& C_M\int_\sigma \frac{p^{\mu}\cdot d\mathbf{\sigma_\mu}}{(2\pi)^3}
	\int_0^1 dx_1 dx_2 \Phi_M(x_1,x_2) \nonumber\\ &&\times f_a({\bf r},x_1\mathbf{p})f_b({\bf r},x_2\mathbf{p}).\
	\label{eq:meson}
\end{eqnarray}
where $C_M$ is the spin degeneracy factor of a given meson species, $\sigma$ is the hypersurface of hadronization, $\varPhi_M$ is the effective wave function squared of mesons, $x_{1,2}$ are light cone coordinates defined as ${\bf p}_{1,2}=x_{1,2}\bf p$, and $f_{a,b}$ are the parton phase-space distributions. The $\varPhi_M$ is parameterized as Gaussian type $\Phi_M\sim \exp (-\frac{(x_1-x_a)^2+(x_2-x_b)^2}{\sigma_M^2})\delta(x_1+x_2-1)$, where $\sigma_M$ is the variance, $x_{a,b}=m_{1,2}/(m_1+m_2)$ are the peak values, and $m_{1,2}$ are the masses of the constituent partons. Similar expression can be derived for baryons.

The quark phase-space distribution is parameterized in a viscous blastwave~\cite{Yang:2016rnw,Yang:2020oig,Yang:2022yxa}, based on the Retiere and Lisa blastwave \cite{Retiere:2003kf}. The quark distribution is given by
\begin{equation}
	f(r,p) = f_{0}(r,p) + \delta f_{\rm {shear}}(r,p)+\delta f_{\rm{bulk}}(r,p)
	\label{eq:fdist}
\end{equation}
where $f_0(r,p) = 1/(e^{(u^\mu p_{\mu}-\mu_i)/T}\mp 1)$ is the equilibrium Bose/Fermi distribution as a function of the flow field $u^\mu$, particle momentum $p^\mu$, the local temperature $T$ and chemical potentials $\mu_i=b_i\mu_B + s_i\mu_S + q_i\mu_Q$ with baryon number $b_i$, strangeness $s_i$ and electric charge $q_i$ (we assume $\mu_Q=0$ in this work). $\delta f_{\rm {shear}}$ and $\delta f_{\rm{bulk}}$ denote corrections from the shear and bulk viscosities, respectively. For the shear viscous corrections, we use the Grad's method \cite{Song:2007ux,Damodaran:2020qxx}
\begin{equation}
	\delta f_{\rm {shear}} = \frac{1}{2T^2} \frac{p_\mu p_\nu}{\epsilon+P}\pi^{\mu\nu} f_{0}(1\pm f_{0})
	\label{eq:shear}
\end{equation}
where $\epsilon$ is the energy density, $P$ is the pressure, $\pi^{\mu\nu}$ is the shear stress tensor and $+(-)$ for bosons (fermions). In the Navier-Stokes approximation $\pi^{\mu\nu}=2\eta \sigma^{\mu\nu}$ where $\eta$ is the shear viscosity and $\sigma^{\mu\nu}$ is the shear gradient tensor defined as $
\sigma^{\mu\nu} = \frac{1}{2} \left( \nabla^\mu u^\nu + \nabla^\nu u^\mu \right) - \frac{1}{3} \Delta^{\mu\nu} \nabla_\lambda u^\lambda$ with flow field $u^\mu$, $\nabla^\mu = \Delta^{\mu\nu} \partial_\nu$ and $\Delta^{\mu\nu} = g^{\mu\nu} - u^\mu u^\nu$. The spatial derivatives in  $\sigma^{\mu\nu}$ can be obtained directly. The time derivatives cannot be given by blastwave itself, and are determined by solving ideal hydrodynamics~\cite{Yang:2022yxa}.

For the bulk viscous corrections, we use the 14-moment approximation \cite{Denicol:2014vaa,Paquet:2015lta}
\begin{equation}
	\delta f_{\rm bulk}=- f_0(1\pm f_0) \Pi \frac{\tau_\Pi}{\zeta} \left[ \frac{1}{3} \frac{m^2}{T} \frac{1}{p^\mu u_\mu}+\frac{p^\mu u_\mu}{T} \left(c_s^2-\frac{1}{3} \right) \right]
	\label{eq:bulk1}
\end{equation}
where $\zeta$ is the bulk viscosity, $\Pi$ is the bulk viscous pressure, $\tau_\Pi$ is the bulk relaxation time and  $c_{s}^2$ is the velocity of sound squared. At the first order approximation, one has $\Pi=-\zeta \partial_\mu u^\mu$.  Following Ref.\ \cite{Denicol:2014vaa}, the ratio of bulk viscosity to bulk relaxation time is given by
\begin{equation}
	\frac{\zeta }{\tau _{\Pi }}=\left( \frac{1}{3}
	-c_{s}^{2}\right) \left( \varepsilon +P \right) -\frac{2}{9}\left(
	\varepsilon-3P\right) -\frac{m^{4}}{9}I_{-2,0}
\end{equation}
with $I_{-2,0}$ defined as $I_{-2,0}=\frac{g}{(2\pi)^3}\int \frac{d^3 \mathbf{p}}{E} \frac{f_{0}(r,p)}{ (p^{\mu}  u_{\mu})^2}$ and $g$ the degeneracy factor. We note that the fluidity of a system at finite chemical potential should be evaluated by the ratio of shear viscosity over the enthalpy multiplied by the temperature, $\eta T/(\epsilon + P)$ \cite{Liao:2009gb}, where $\epsilon + P=Ts+\mu_B n_B$. When $\mu_B=0$, this quantity reduces to $\eta/s$.

Let us now describe the blastwave parameterization for the flow field $u^\mu$. Here $R_{x,y}$ are the semi axes of the fireball at freeze-out, $\rho=\sqrt{x^2/R_x^2+y^2/R_y^2}$ is the reduced radius, $\eta_s = \frac{1}{2}\ln\frac{t+z}{t-z}$ is the space-time rapidity. The hypersurface is assumed to be constant $\tau=\sqrt{t^2-z^2}$. The flow field is parameterized as
\begin{eqnarray}
	u^{\mu}&=&(\cosh\eta_s\cosh\eta_T,\sinh\eta_T\cos\phi_b, \nonumber\\ && \qquad \sinh\eta_T\sin\phi_b,\sinh\eta_s\cosh\eta_T)
	\label{eq:flow}
\end{eqnarray} where $\eta_T$ is the transverse flow rapidity and $\phi_b$ is the azimuthal angle of $u^\mu$ in the transverse plane. $\eta_T$ is given by the transverse velocity $v_T=\tanh\eta_T$ with
\begin{equation}
	v_T=\rho^n (\alpha_0+\alpha_2\cos2\phi_b)
\end{equation}
where $\alpha_0$ is the average surface velocity, $\alpha_2$ is an elliptic deformation of the flow field and $n$ is a power term. In this work, we use a linear expression and set $n=1$. The transverse flow vector is chosen to be perpendicular to the elliptic surface at $\rho=1$. The ratio $R_y/R_x$ significantly influences elliptic flow, so we choose $R_y/R_x$ as a fit parameter and constrain $R_x$, $R_y$ and $\tau$ by adding a simple geometric estimate $R_x \approx (R_0-b/2)+ 0.65 \tau (\alpha_0+\alpha_2)$ where $R_0$ is the radius of the colliding nucleus and $b$ is the impact parameter. The values of $b$ used for each centrality bin are based on Glauber Monte Carlo simulations for related experiments \cite{ALICE:2013hur,HADES:2017def}.

It should be noted that the viscous blastwave offers a simplified representation of the flow field and freeze-out hypersurface, serving as an approximate snapshot of a viscous hydrodynamic system at a fixed time~\cite{Teaney:2003kp,Jaiswal:2015saa,Yang:2016rnw,Yang:2020oig,Yang:2022yxa}. 
The dissipative effects
before freeze-out are effectively incorporated into the parameterized flow field $u^{\mu}$. Consequently, the viscous blastwave carries information on the viscosities of the fluid at a specific time, e.g., the QGP at hadronization in our present work.

\begin{table}[tb]
	\caption{\label{tab:data} Experimental data of $\phi$ and $\Omega$ used in our analysis. Values in parentheses are centralities of $\Omega$ when they are different from $\phi$.}
	\centering
	\resizebox{\linewidth}{!}{
		\begin{tabular}{l|l|l}
			\hline
			$\sqrt{s_{NN}}$(GeV) & \multicolumn{2}{c}{Centrality}  \\
			\hline
			\multirow{2}*{19.6,39} & $v_2(p_T)$ &0-10\%, 10-40\%, 40-80\% \cite{STAR:2015rxv,Liu:2022mxa,Dixit:2022geb} \\
			& $\frac{dN}{d^2p_T dy}$ &0-10\%, 20-30\%(20-40\%), 40-60\% \cite{STAR:2015vvs,STAR:2019bjj} \\
			\hline
			\multirow{2}*{54.4,62.4} & $v_2(p_T)$  &0-10\%, 10-40\%, 40-80\% \cite{STAR:2015rxv,STAR:2022tfp}\\
			& $\frac{dN}{d^2p_T dy}$ &0-20\%, 20-40\%, 40-60\% \cite{STAR:2008bgi,STAR:2010yyv}\\
			\hline
			\multirow{2}*{200} & $v_2(p_T)$ &0-30\%, 30-80\% \cite{STAR:2015gge} \\
			& $\frac{dN}{d^2p_T dy}$ & 10-20\%(0-5\%), 40-50\%(40-60\%) \cite{STAR:2006egk,STAR:2007mum} \\
			\hline
			\multirow{3}*{2760} & $v_2(p_T)$ 	&\multirow{2}*{10-20\%, 20-30\%, 30-40\%,	40-50\%  \cite{ALICE:2014wao,ALICE:2017ban}}  \\
			& $\frac{dN}{d^2p_T dy}(\phi)$                      \\
			& $\frac{dN}{d^2p_T dy} (\Omega)$ & 10-20\%, 20-40\%, 40-60\% \cite{ALICE:2013xmt} \\
			\hline
	\end{tabular}	}
\end{table}

\subsection{Experimental data and fit parameters}
We utilize the transverse-momentum~($p_T$) spectra and elliptic flows $v_2$ of $\phi$ and $\Omega$ as our observables. The data we used are from the STAR collaboration, covering Au+Au collisions at 19.6, 39, 54.4, 62.4 and 200 GeV~\cite{STAR:2006egk,STAR:2007mum,STAR:2008bgi,STAR:2010yyv,STAR:2015gge,STAR:2015rxv,STAR:2015vvs,STAR:2019bjj,STAR:2022tfp,Liu:2022mxa,Dixit:2022geb} (STAR Preliminary data for $v_2$ at 19.6 GeV), and the ALICE collaboration for Pb+Pb collisions at $\sqrt{s_{NN}}$=2.76 TeV~\cite{ALICE:2013xmt,ALICE:2014wao,ALICE:2017ban} around mid-rapidity, as listed in Tab.\ \ref{tab:data}. Due to data availability, the centrality bins for spectra are slightly different from that of $v_2$. For Au+Au at 19.6, 39, 54.4 and 62.4 GeV, we use data of $\Omega^-$. For Au+Au at 200 GeV and Pb+Pb at 2.76 TeV, we use data of $\Omega^-+\bar{\Omega}^+$. For Au+Au collisions at $\sqrt{s_{NN}}=$ 7.7,11.5 and 27 GeV, the current data of $\phi$ and $\Omega$ have very few points and large uncertainties, so we will not use them in our analysis. As for the spectra of $\phi$ and $\Omega$ at $\sqrt{s_{NN}}=$ 54.4 GeV, there is currently no available data and we use the spectra from $\sqrt{s_{NN}}=$ 62.4 GeV instead.

For each collision energy, we perform a combined analysis of the available centrality bins. The fitted $p_T$ ranges for $\phi$ and $\Omega$ are given in Tab.\ \Ref{tab:range}. We have checked that the correction $\delta f(r,p)$ is small for both $\phi$ and $\Omega$ with the above fit ranges, i.e., less than $20\%$ of $f_0$ for the majority of points (very few points going up to $40\%$ of $f_0$), which is much smaller than the commonly adopted upper bound $\sim 1$~\cite{JETSCAPE:2020mzn,Teaney:2003kp}, ensuring the applicability of the viscous corrections.

The temperature $T$ and baryon chemical potentials $\mu_B$ at hadronization are taken from Ref.\ \cite{Andronic:2017pug}, by assuming their values are close to the corresponding values at chemical freeze-out for different collision energies. Besides, the strangeness chemical potentials $\mu_S$ are obtained by extrapolating the results from Ref.\ \cite{STAR:2019bjj}. The values of $(T, \mu_B,\mu_S)$ used in our calculation for different collision energies are listed in Tab.\ \ref{tab:range}. To model the yields of $\phi$ and $\Omega$, we introduce a fugacity factor $\gamma_{s,\bar{s}}$, setting $\gamma_s=0.65$ for 19.6-62.4 GeV and $\gamma_s=0.8$ for 200, 2760 GeV. Note that the value of $\gamma_s$ for 19.6-62.4 GeV are smaller than those obtained in thermal models~\cite{STAR:2010yyv,STAR:2017sal}. The value of $\gamma_s$ for 200, 2760 GeV is derived from quark coalescence model~\cite{Pu:2018eei}. In order to assess the impact of $\gamma_s$, we increase the value of $\gamma_s$ by 25\%, setting $\gamma_s=0.81$ for 19.6-62.4 GeV and $\gamma_s=1$ for 200-2760 GeV, and we observe minor effects on the final results. For instance, with $\gamma_s=0.81$ for Au+Au collisions at 62.4 GeV, the value of $T/(\epsilon + P)$ decreases by only 7\% and $\zeta T/(\epsilon + P)$ decreases by a mere 3\% when compared to default results.
	
Regarding other constants, we specify the hadron wave function variances as $\sigma_M=0.3$ and $\sigma_B=0.1$. We have confirmed that the variations in the wave function's variances have minimal impact on our results. We use sound speed squared $c^2_s=0.15$ (see Eq.\ (\ref{eq:bulk1})) for the QGP at hadronization \cite{Huovinen:2009yb,Monnai:2019hkn}, quark mass $m_s=500$ MeV, spin degeneracy factor $C_M=3$ for $\mathrm{\phi}$ and $C_B=4$ for $\Omega^-$.

\begin{table}[tb]
	\caption{\label{tab:range} Values of temperature ($T$), baryon chemical potential ($\mu_B$), strangeness chemical potential ($\mu_S$) and fit ranges for $\phi$ and $\Omega$ observables used in quark recombination for different collision energies.}
	\centering
	\begin{tabular}{l|l|llllll} 
		\hline
		\multicolumn{2}{c|}{$\sqrt{s_{NN}}$ (GeV)}  &19.6  & 39 &54.4 &62.4 & 200 & 2760\\
		\hline
		\multicolumn{2}{c|}{$T$ (MeV) }  & 155  & 157.5 & 158.5 & 158.5 & 160 & 160 \\
		\hline
		\multicolumn{2}{c|}{$\mu_B$ (MeV) } & 197  & 107  & 79 & 69 & 22  & 0 \\
		\hline
		\multicolumn{2}{c|}{	$\mu_S$ (MeV)} & 46   & 29 & 21  & 18 & 8 & 0 \\
		\hline
		\multirow{2}*{$p_T$-range (GeV/$c$)}&$\phi$ & $<2.3$  & $<2.4$ & $<2.6$  & $<2.7$ & $<2.6$ & $<2.8$ \\
		\cline{2-8}
		&$\Omega$ & $<3.2$   & $<3.0$ & $<3.9$  & $<3.8$ & $<4.0$ & $<3.8$ \\
		\hline
	\end{tabular}	
\end{table}

The parameters left in our model are ($\tau$, $\alpha_0$, $\alpha_2$, $R_y/R_x$, $\eta T/(\epsilon + P)$, $\zeta T/(\epsilon + P)$), which can be determined by Bayesian analysis of the experimental data. For each centrality bin at each collision energy, the fluid has unique values for ($\tau$, $\alpha_0$, $\alpha_2$, $R_y/R_x$) and shared values for ($\eta T/(\epsilon + P)$, $\zeta T/(\epsilon + P)$) and thus we have 3$\times$4+2=14 parameters for $\sqrt{s_{NN}}$= 19.6-62.4 GeV, 2$\times$4+2=10 parameters for $\sqrt{s_{NN}}$= 200 GeV and 4$\times$4+2=18 parameters for $\sqrt{s_{NN}}$= 2.76 TeV.

\subsection{Bayesian method}
To determine the above parameters, we employ the Bayesian analysis package from the Models and Data Analysis Initiative (MADAI) project \cite{MADAI:2013,Bernhard:2016tnd}. The MADAI package includes a Gaussian process emulator and a Bayesian analysis tool. According to Bayes' theorem, for model parameters $\rm x=(x_1,x_2,x_3,...)$ and experimental observables $\rm y=(y_1,y_2,y_3,...)$, the probability for the true parameters $\rm x_\star$ is
	\begin{equation}
		P({\rm x}_\star|X,Y,{\rm y}_{\rm {exp}}) \propto P(X,Y,{\rm y}_{\rm {exp}}|{\rm x}_\star) P({\rm x}_\star).
		\label{eq:bayes}
	\end{equation}
	The left-hand side is the \emph{posterior} probability of ${\rm x}_\star$ given the design ($X$, $Y$) and the experimental data ${\rm y}_{\rm {exp}}$. On the right-hand side, $P({\rm x}_\star)$ is the \emph{prior} probability and $P(X,Y,{\rm y}_{\rm {exp}}|{\rm x}_\star)$ is the likelihood, i.\ e.\ , the probability of the model describing the data ${\rm y}_{\rm {exp}}$ at ${\rm x}_\star$, given by
	\begin{equation}
		P(X,Y,{\rm y}_{\rm {exp}}|{\rm x}_\star)\propto\exp (
		-\frac{1}{2} \Delta {\rm y}^\top \Lambda^{-1} \Delta {\rm y})
	\end{equation}
	where $\Delta {\rm y}={\rm y}_\star - {\rm y}_{\rm {exp}}$ is the difference between the measurement and the prediction, and $\Lambda$ is the covariance matrix including the experimental and model uncertainties. Further information is available in Refs.\ \cite{Bernhard:2016tnd,JETSCAPE:2020mzn}.

\begin{figure}[tbh]
	\centering
	\begin{minipage}{0.235 \textwidth}
		\includegraphics[width=1\textwidth]{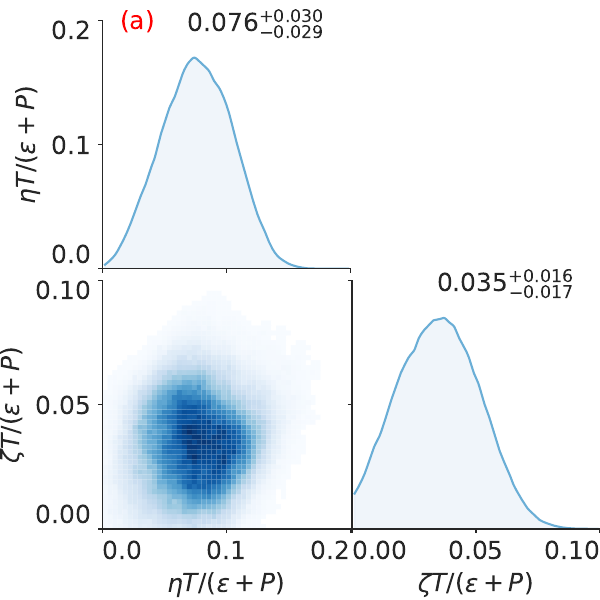}
	\end{minipage}
	\begin{minipage}{0.235 \textwidth}
		\includegraphics[width=1\textwidth]{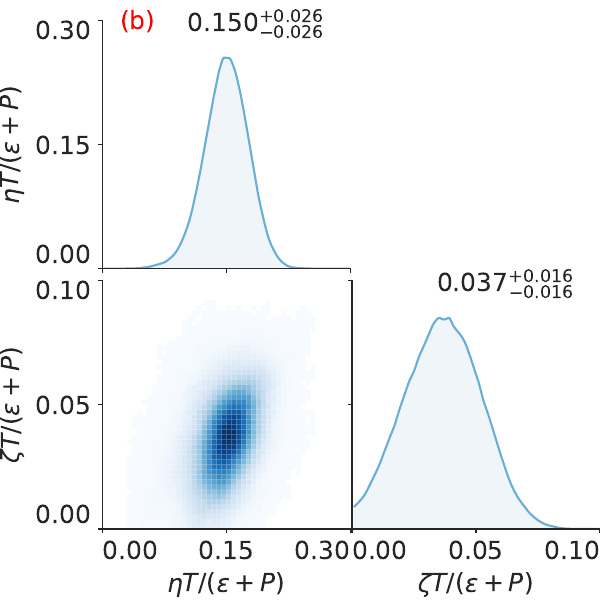}
	\end{minipage}
	\caption{\label{fig:posterior}
		Posterior distributions of $\eta T/(\epsilon + P)$ and $\zeta T/(\epsilon + P)$ for Au+Au 19.6 GeV (a) and 62.4 GeV (b). The numbers indicate the median values with the 68.3\%-credibility range.}
\end{figure}

We adopt a uniform prior distribution for all model parameters. For instance, for 10-40\% centrality bin at $\sqrt{s_{NN}}$= 19.6 GeV, we set prior ranges 5.8-7.8 fm/$c$ for $\tau$, 0.44-0.6$c$ for $\alpha_0$, 0.012-0.042$c$ for $\alpha_2$, 1.12-1.32 for $R_y/R_x$ and obtain posterior values ($\tau, \alpha_0, \alpha_2, R_y/R_x$)=(6.8 fm/$c$, 0.53$c$, 0.03$c$,1.24). Besides, we set prior ranges 0-0.2 for $\eta T/(\epsilon + P)$, 0-0.12 for $\zeta T/(\epsilon + P)$ at 19.6 GeV and obtain $\eta T/(\epsilon + P)=0.076$ and $\zeta \rm {T/(\epsilon + P)}= 0.035$. The same procedure is applied to other centralities and energies.

After setting prior ranges for each parameter, we generate a set of training points within the parameter space and calculate all fitted observables at each training point. The MADAI package then builds a Gaussian process emulator, which can estimate the observables for random parameter values. Finally a Markov Chain Monte Carlo (MCMC) provides a likelihood analysis and gives the maximum likelihood or best-fit parameters. Here for each collision energy we use N=500 training points. To validate proper functioning of the Bayesian analysis, we perform a closure test and confirm the Bayesian framework correctly reproduces model parameters within reasonable uncertainties. The likelihood analysis has used ${\rm N}_\star=2\times 10^6$ predicted points to search for the best-fit parameters, which is sufficient for MCMC to converge.

\section{Results and discussions}
\label{sec:results}

Using the data and parameters discussed earlier, we perform a model-to-data comparison with MADAI package and obtain the best-fit parameters, which are defined as the mean value given by the maximum likelihood analysis. 
As an example, we illustrate in Fig.\ \ref{fig:posterior} the univariate posterior distributions of $\eta$ and $\zeta$ for Au+Au collisions at $\sqrt{s_{NN}}$=19.6 GeV (a) and 62.4 GeV (b). For 19.6 GeV, we obtain $\eta T/(\epsilon + P)=0.076^{+0.030}_{-0.029}$ and $\zeta \rm {T/(\epsilon + P)}= 0.035^{+0.017}_{-0.016}$. For 62.4 GeV, we obtain $\eta T/(\epsilon + P)=0.15\pm 0.026$ and $\zeta T/(\epsilon + P)= 0.037\pm 0.016$, both at a $68.3\%$ confidence level~(C.L.). Similar results can be obtained for other collision energies.

\begin{figure}[tb]
	\centering
	\includegraphics[width=0.48\textwidth,height=0.5\textwidth]{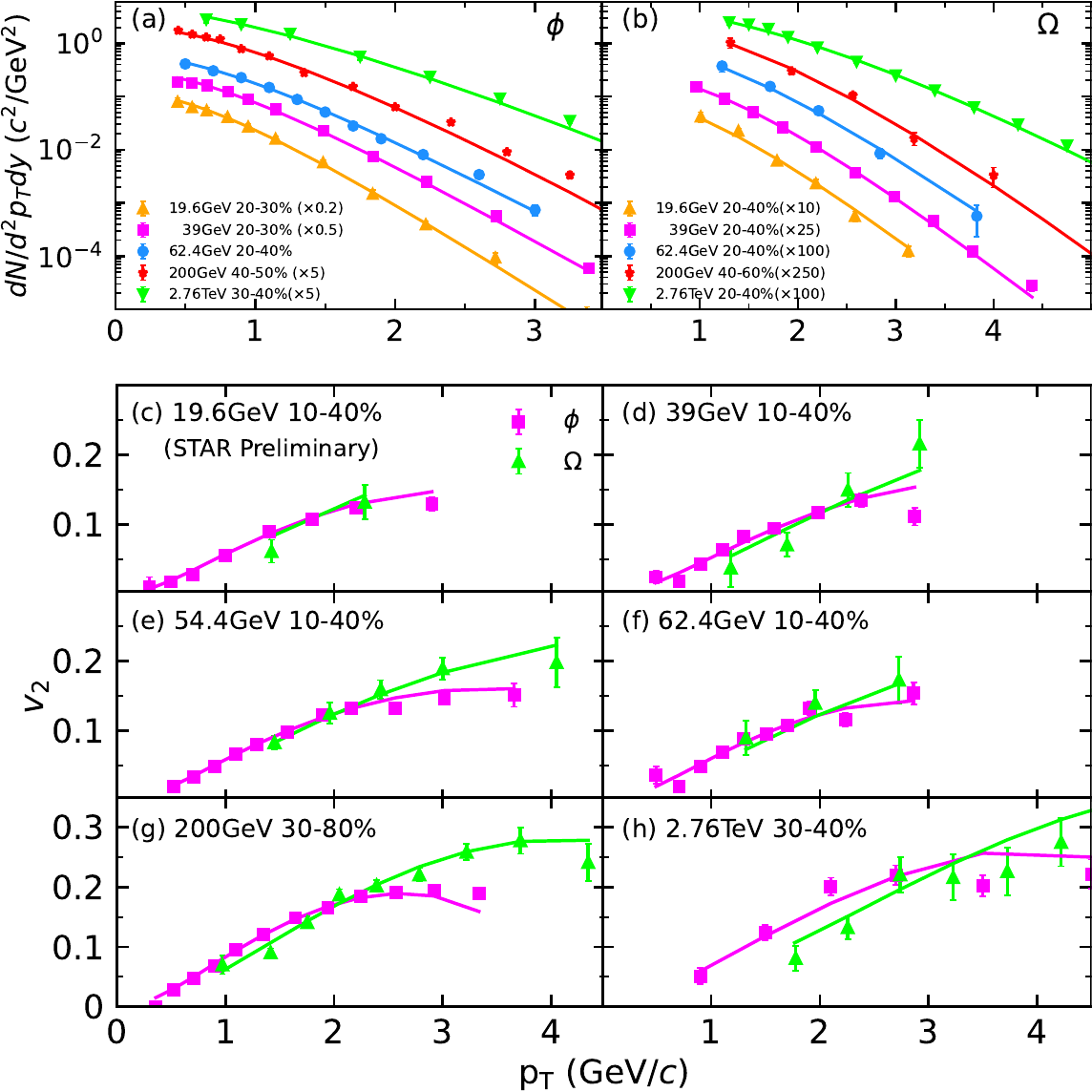}
	\caption{\label{fig:spectrav2}
		Transverse-momentum spectra and elliptic flows $v_2$ of $\phi$ mesons and $\Omega$ baryons (selected centrality bins). For Au+Au at 19.6-62.4 GeV, we use data of $\Omega^-$. For Au+Au at 200 GeV and Pb+Pb at 2.76 TeV, we use data of $\Omega^-+\bar{\Omega}^+$. Solid lines are recombination calculation using the best-fit parameters. The data from STAR~\cite{STAR:2006egk,STAR:2007mum,STAR:2008bgi,STAR:2010yyv,STAR:2015gge,STAR:2015rxv,STAR:2015vvs,STAR:2019bjj,STAR:2022tfp,Liu:2022mxa,Dixit:2022geb} and ALICE~\cite{ALICE:2013xmt,ALICE:2014wao,ALICE:2017ban} are included for comparison.}	
\end{figure}

With the best-fit parameters provided by Bayesian analysis, we can calculate transverse momentum spectra and elliptic flows of $\phi$ and $\Omega$ and compare with experimental data. Fig.\ \ref{fig:spectrav2} shows our theoretical predictions and the corresponding experimental data for $p_T$ spectra and $v_2$ of $\phi$ and $\Omega$ in selected centralities at different collision energies. As seen from Fig.\ \ref{fig:spectrav2}, our calculations describe data rather well.

\begin{figure}[tb]
	\centering
	\includegraphics[width=0.42\textwidth]{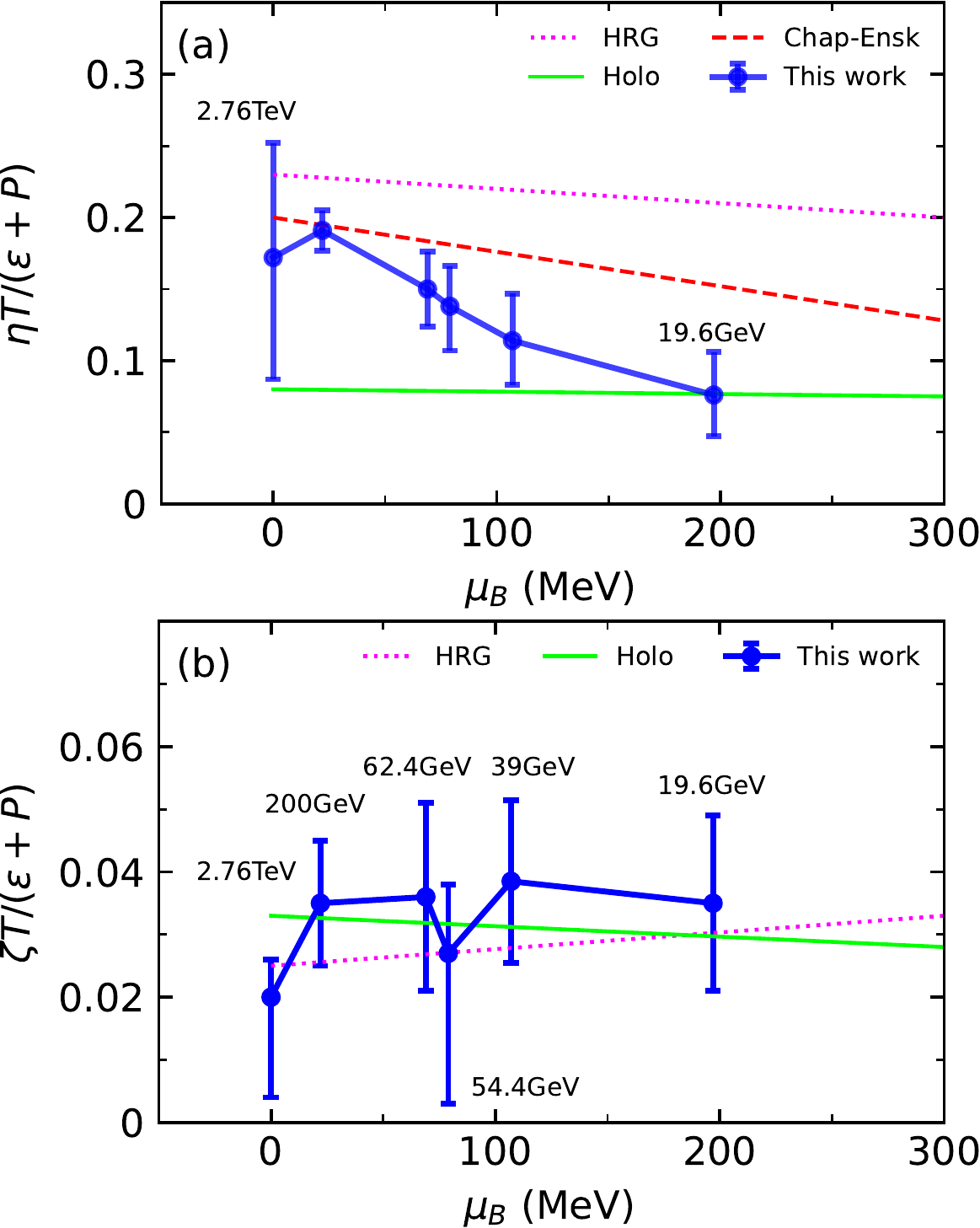}
	\caption{\label{fig:result}Baryon chemical potential dependence of shear (a) and bulk (b) viscosities for QGP/hadronic matter at hadronization with $T\sim 160$ MeV(see text for details).}
\end{figure}

Figure\ \ref{fig:result} shows our Bayesian inference of the $\eta T/(\epsilon + P)$ (a) and $\zeta T/(\epsilon + P)$ (b) at 68.3\% C.L. for the QGP at hadronization or $T\sim 160$ MeV with different $\mu_B$. The range of baryon chemical potentials is between $\mu_B=0$ and $\mu_B=200$ MeV, which corresponds to collison energy varying from 2.76 TeV (most left) to 19.6 GeV (most right) \cite{Andronic:2017pug}. For the purpose of comparison, in Fig.\ \ref{fig:result} we also include results from other approaches, i.e., Chapman-Enskog theory (Chap-Ensk)\cite{Denicol:2013nua} for $\eta$, hadron resonance gas model (HRG) \cite{Kadam:2015xsa} and holographic model (Holo) \cite{Grefa:2022sav} for both $\eta$ and $\zeta$. In these approaches, $\eta$ and $\zeta$ are caculated as a function of temperature with different $\mu_B$. Here we take their values at $T = 160$ MeV with $\mu_B=0$ and $\mu_B=300$ MeV ($T = 150$ MeV with $\mu_B=0$ and $\mu_B=500$ MeV for Chap-Ensk). It should be noted that the results from Chap-Ensk and HRG are for hadronic matter.

One sees from Fig.\ \ref{fig:result}(a) that the shear viscosity has a significant dependency on baryon chemical potential and decreases with $\mu_B$, which suggests that the QGP with finite baryon density is more close to perfect fluid than the baryon-free QGP. This trend of shear viscosity, i.e., $\eta T/(\epsilon + P)$ decreases as $\mu_B$ increases, aligns with observations in Chap-Ensk~\cite{Denicol:2013nua}, HRG~\cite{Kadam:2015xsa,McLaughlin:2021dph} and holographic model~\cite{Grefa:2022sav}, but seems to be at variance with the findings in hybrid models \cite{Karpenko:2015xea,Auvinen:2017fjw,Shen:2023awv}.
We note that in the hybrid models, a constant~$\eta/s$~\cite{Karpenko:2015xea,Auvinen:2017fjw} or temperature-independent $\eta T/(\epsilon + P)$~\cite{Shen:2023awv} is assumed for each collision energy and thus the obtained results are temperature-averaged by neglecting the effects of varying $T$ on viscosities of the QGP during the dynamical evolution. This is in contrast with our present results that are for QGP at hadronization with a temperature of $T\sim 160$ MeV.

It should be emphasized that the shear viscosity may have complicated dependence on the temperature and chemical potential (see, e.g., Ref.~\cite{McLaughlin:2021dph}).
As found in Refs.\ \cite{Denicol:2013nua,Kadam:2015xsa}, the decrease of $\eta T/(\epsilon + P)$ with $\mu_B$ is primarily attributed to the rapid increase in entropy density $s$ and very slow increase in $\eta$ with $\mu_B$. In addition, we would like to mention that at $\mu_B=0$, the most recent Bayesian statistical analysis leads to $\eta/s \approx 0.16$ at $T = 160$ MeV~\cite{Nijs:2022rme,Heffernan:2023utr}, consistent with our present result.

On the other hand, Fig.\ \ref{fig:result}(b) indicates rather small $\zeta$ at hadronization for $\mu_B<200$ MeV. In particular, we find $\zeta T/(\epsilon + P)$ is essentially constant within uncertainty with $\zeta T/(\epsilon + P)=0.02\sim 0.04$ for $\mu_B<200$ MeV. Interestingly, our results are quantitatively consistent with HRG and holographic model. However, due to the limited accuracy of our results, we cannot draw any conclusions about the trend for $\mu_B$ dependence of $\zeta$. We note that HRG predicts an increase of $\zeta T/(\epsilon + P)$ with $\mu_B$, while the holographic model predicts an opposite behavior.

Addressing uncertainties in our analysis, we categorize them into two main groups: (a) uncertainties arising from assumptions made in the blastwave parameterization, such as the simplistic ansatz for the flow field and the recombine hypersurface, and (b) uncertainties stemming from errors in experimental data and the quality of the Gaussian emulator. To quantify uncertainty (a), a comparison between hydro simulations and blastwave was conducted in~\cite{Yang:2020oig} for $\mu_B=0$ and found blastwave represents the shear stress in hydro simulations quite well, with uncertainties around 0.02 for $\eta/s$. A future exploration involving a comparison between blastwave and hydro simulations at finite $\mu_B$ is planned. Uncertainty (b) is addressed by the MADAI code and is presented in our final results.

\section{Conclusions}
\label{sec:conclusion}

We have performed Bayesian inference of the shear and bulk viscosities of the QGP at hadronization ($T\sim 160$ MeV) by analyzing the $\phi$ and $\Omega$ data in Au+Au collisions at $\sqrt{s_{\rm NN}}=$ 19.6-200~GeV and Pb+Pb collisions at $\sqrt{s_{\rm NN}}=$ 2.76~TeV, based on the quark recombination model coupled with a viscous blastwave. We find that the specific shear viscosity $\eta T/(\epsilon +P)$ of the QGP at hadronization decreases as $\mu_B$ increases, while the corresponding specific bulk viscosity is essentially constant for $\mu_B<200$ MeV, suggesting that the QGP at $T\sim 160$ MeV with finite baryon density is more close to perfect fluid than that with zero baryon density.

Our work provides valuable reference for future theoretical calculations as well as parameterization of viscosities in hydro simulations for baryon-rich QGP.
Our work is also useful for exploring the dynamics of binary neutron star mergers, given that quark matter is expected to exist in the cores of massive neutron stars and a transition from nuclear to quark matter may happen~\cite{Most:2018eaw}. Furthermore, dramatic changes in transport properties may be considered as a signal of phase transition, and a thorough understanding of the baryon density dependence of viscosities of QCD matter is of great significance for the exploration of phase transition~\cite{McLaughlin:2021dph}. Our study represents a step forward in achieving this objective, especially if more high quality $\phi$ and $\Omega$ data in Au+Au collisions at $\sqrt{s_{NN}}=$ 7.7 and 11.5 GeV or even lower collision energies would be provided at RHIC in future.

\begin{acknowledgments}
	
The authors would like to thank Rainer J. Fries, Defu Hou, Feng Li, Jie Pu and Kai-Jia Sun for useful discussions. This work was supported in part by the National Natural Science Foundation of China under Grant Nos. 12205182, 12235010, 11625521 and 12375124, the National SKA Program of China No. 2020SKA0120300, and the Science and Technology
Commission of Shanghai Municipality under Grant No. 23JC1402700. Y.S. thanks the sponsorship from Yangyang Development Fund.

\end{acknowledgments}

\bibliography{ref}

\end{document}